\documentclass[journal=achemso-jpcafh,manuscript=article,layout=twocolumn]{achemso}

\usepackage[version=3]{mhchem} 
\usepackage{algorithm2e}
\usepackage{algpseudocode}
\usepackage[compatibility=false]{caption}
\usepackage{multirow}
\usepackage{textcomp}
\usepackage{widetext}
\usepackage{amsmath}
\usepackage{cases}
\usepackage{graphicx}
\usepackage{enumitem}
\usepackage{tabularx}
\usepackage{float}
\usepackage{color}
\usepackage{rotating}

\author{Bo Peng}
\email{peng398@pnnl.gov}
\affiliation{Physical and Computational Science Division,
Pacific Northwest National Laboratory, Richland, WA 99352, USA}
\phone{(509) 371-7578}
\author{Ajay Panyala}
\email{ajay.panyala@pnnl.gov}
\affiliation{Advanced Computing, Mathematics, and Data Division,
Pacific Northwest National Laboratory, Richland, WA 99352, USA}
\author{Karol Kowalski}
\email{karol.kowalski@pnnl.gov}
\affiliation{Physical and Computational Science Division,
Pacific Northwest National Laboratory, Richland, WA 99352, USA}
\author{Sriram Krishnamoorthy}
\email{sriram@pnnl.gov}
\affiliation{High-Performance Computing,
Pacific Northwest National Laboratory, Richland, WA 99352, USA}

\title[gfcclib]
  {GFCCLib: Scalable and Efficient Coupled-Cluster Green's Function Library for Accurately Tackling Many-Body Electronic Structure Problems}

\begin{document}
\begin{abstract}
Coupled-cluster Green's function (GFCC) calculation has drawn much attention in the recent years for targeting the molecular and material electronic structure problems from a many-body perspective in a systematically improvable way. However, GFCC calculations on scientific computing clusters usually suffer from expensive higher dimensional tensor contractions in the complex space, expensive inter-process communication, and severe load imbalance, which limits it's routine use for tackling electronic structure problems. Here we present a numerical library prototype that is specifically designed for large-scale GFCC calculations. The design of the library is focused on a systematically optimal computing strategy to improve its scalability and efficiency. The performance of the library is demonstrated by the relevant profiling analysis of running GFCC calculations on remote giant computing clusters. The capability of the library is highlighted by computing a wide near valence band of a fullerene C60 molecule for the first time at the GFCCSD level that shows excellent agreement with the experimental spectrum. 
\end{abstract}


\section{Program summary}
\noindent
\textbf{Program Title:} GFCCLib

\noindent
\textbf{Program Files doi:}  10.24433/CO.5131827.v1

\noindent
\textbf{Licensing provisions:} MIT License

\noindent
\textbf{Programming language:} C++

\noindent
\textbf{Nature of problem:} The applications of coupled cluster Green's function on large scale molecular electronic structure problems suffer from expensive higher dimensional tensor contractions in the complex space, expensive inter-process communication, and severe load imbalance. Tackling these issues are a key step in building high-performance coupled cluster Green's function library for its routine use in large scale molecular science.

\noindent
\textbf{Solution method:} We have developed a C++ library for large scale molecular GFCC calculations on high-performance computing clusters. We provide implementations for high dimensional tensor algebra for many-body methods (TAMM), Cholesky decomposition of high dimensional electron repulsion integral tensors, process group technique for mitigating load imbalance. The library is written in C++. The source code, tutorials and documentation are provided online. A continuous integration mechanism is set up to automatically run a series of regression tests and check code coverage when the codebase is updated.

\section{Introduction}
Faster and more accurate description of the electronic structure of large quantum systems has always been stimulating the new development of \textit{ab initio} electron correlation theory \cite{march67, fetter2012quantum, linderberg2004propagators, paldus75_105, cederbaum77_205, joergensen2012second, szabo2012modern, oddershede87propagator, mattuck2012guide, harris2020algebraic, bartlett07_291, shavitt2009many}. 
For example, larger fullerene molecules (e.g. C60 and C70) and their derivatives have been widely applied in the molecular electronics and nanostructured devices, especially in organic photovoltaics (OPVs) \cite{Yu95_1789, brabec01_374, tang86_183}. In comparison with the traditional silicon-based photovoltaic materials, the fullerene-based OPVs have exhibited promising photoelectric properties, and their power conversion efficiency has been gradually improved over the past years (from 1\% to over 9\%) \cite{mayer07_1369,Deibel10_096401}. However, the fundamental understanding of the correlation between the photoelectric properties and the structure of these fullerene-based materials is still not fully clear, in particular how the open-circuit voltage is affected by the electronic structure that is associated with certain structures. 
 
%
Previously, only single particle pictures (for example density-functional theory, DFT) of the electronic structure of fullerene and derivatives have been reported \cite{tiago08_084311, akaike08_023710, zhang08_19158, tiago09_195410, blase11_115103}, which only shows qualitative agreement with experiment. Of course, one can invoke different density functionals in the DFT calculation to test the agreement with the experiment. Unfortunately,  there is no systematic way to improve the single particle DFT results, and separate calculations for different states would be required. On the other hand, one can directly compute the one-particle many-body Green's function (MBGF) to capture the key electronic properties of the ionization and attachment process. Typical approaches include {\it GW} method,\cite{hedin65_a796,faleev04_126406, schilfgaarde06_226402, louie06_216405, louie11_186404,setten13_232} outer-valence Green's function (OVGF) method,\cite{cederbaum75_290, cederbaum84_57, ortiz97, ortiz13_123} and algebraic-diagrammatic construction (ADC)\cite{schirmer82_2395,  cederbaum84_57, dreuw15_82} approximation scheme. 
Both the {\it GW} method and OVGF method rely on the finite many-body perturbative expansion of the self-energy via the Dyson equation, and have been proved by numerous studies of weakly and moderately correlated molecular systems to provide accurate single particle properties. However, when many-body effects become crucial, as often featured by the satellite states in the ionization process out of the inner valence band where poles will appear in the analytical structure of the self-energy,\cite{cederbaum77_L549, cederbaum75_2160, cederbaum80_481} one can not properly describe the Green's function and/or self-energy by a finite expansion. In contrast, infinite partial summations of the self-energy perturbation series, for example the ADC approximation scheme (especially the third-order ADC method, ADC(3)), has been proved to be able to provide qualitative description of the many-body poles in the analytical structures of the self-energy. 

Alternatively, the improved description of the many-body effect in the Green's function calculations might systematically be achieved by incorporating the one-particle MBGF with correlated wave function expansions. For example, the possibility of  utilizing systematically-improvable highly-correlated wave function methodologies as impurity solver to describe local Green's function or corresponding self-energies in dynamical-mean field theories (DMFT) has drawn considerable interest recently.\cite{kotliar96_13, kotliar06_865, vollhardt12_1, millis06_155107, millis06_076405, zgid11_094115, zgid12_165128, zhu2019_115154, zgid19_6010} Among these systematically-improvable many-body approaches,
the Green's function coupled cluster (GFCC) methodology has attracted much attention in recent years for the molecular and material quantum chemical calculations \cite{nooijen92_55,  nooijen93_15, nooijen95_1681, kowalski14_094102,  kowalski16_144101, kowalski16_062512, chan16_235139, hirata17_044108, kowalski18_561,kowalski18_4335, kowalski18_214102, matsushita18_034106, kowalski18_3, berkelbach18_4224, peng19_3185, zgid19_6010, peng20_011101, matsushita20_012330, bauman20}, and could be one of the ideal many-body tools to treat a complex molecular  system  like  C60. Inheriting the merits of many-body Green's function method and coupled cluster method, the GFCC method is able to describe the electron propagation in a systematically improvable many-body way for the molecular and material quantum systems. However, despite these features, the GFCC method scales polynomially with system size, that is if $N$ is the number of basis functions representing the problem size, the scaling is $N^x$ with the value of $x$ relying on the approximation that one takes in the calculation. Usually, cruder approximation that requires smaller $x$ to make the calculations cheaper will also lead to larger error. Thus, similar to the relevant coupled cluster theory
\cite{cizek66_4256,paldus72_50,purvis82_1910,paldus07,bartlett07_291}, 
the GFCC method forms a hierarchy in terms of computational efficiency and accuracy. As a rule of thumb, for small and medium quantum systems that are described by $<$500 basis functions, one may comfortably use GFCC with singles and doubles (GFCCSD, an $N^6$ method) to compute the corresponding many-body electronic structures. For relatively large quantum systems that are described by $>$500 basis functions, the GFCC calculations on conventional computing clusters can only compute a small number of ionized states, which would make the method less predictable if important states were not computed.

For example, the GFCC matrix was used to be obtained through diagonalizing the non-Hermitian equation-of-motion coupled cluster (EOM-CC) Hamiltonian matrix in the ($N \pm 1$)-particle space to  construct the sum-over-states representation. The performance of solving such an eigen-problem, 
if the EOM-CC Hamiltonian is a dense matrix,
will significantly deteriorate if the dimension of the Hamiltonian grows over $10^{10}$. Not to mention that only limited number of states can be obtained with the help of iterative methods. 
To reduce the computational cost of solving the eigen-problem in the EOM step, as well as the cost in the preceding ground state CC step, early attempts have mainly focused on perturbative truncation of the similarity transformed Hamiltonian to achieve an $\mathcal{O}(N^5)$ scaling \cite{stanton95_1064}. More recently, various local descriptions of the correlation wave function have emerged to facilitate the development of reduced-scaling coupled-cluster methods. In particular the pair natural orbitals (PNOs) introduced almost half century ago \cite{meyer73_1017,edmiston66_1833,edmiston68_192,ahlrichs75_275} have been resurrected and further developed by Neese and co-workers as domain-based local pair natural orbitals (DLPNO) \cite{neese13_034106,neese16_024109} to combine with the EOM-CC method to reduce the size of the space for the diagonalization \cite{neese16_034102,neese18_244101,neese19_164123}.
Alternatively, the GFCC matrix can be obtained directly by solving a set of shifted linear systems involving the EOM-CC Hamiltonian at a frequency of interest. Therefore, due to its algebraic structure, systems of linear equations can be solved simultaneously over multiple frequencies that is well suited for massively distributed computing architectures. Using similar methods, pilot calculations have been previously reported for uniform electron gas \cite{chan16_235139}, light atoms \cite{matsushita18_034106}, heavy metal atoms \cite{matsushita18_224103}, and simple 1-D periodic systems \cite{matsushita18_204109}. More recently, the GFCC as an impurity solver has been applied in the embedding computing framework to compute the electronic structure of complex materials.\cite{zhu2019_115154,zgid19_6010}. However, large GFCC calculations (with $N>$ 1000) have never been tried until recently, where the valence band electronic structures of a series of large DNA fragments have been computed using GFCC approach showing the possibility and potential of this approach in the large scale applications \cite{peng20_011101}. 

Remarkably, the algebraic structure of the  GFCC equations  enables  highly scalable implementations utilizing multiple levels of parallelism.
To further optimize the GFCC infrastructure, three bottlenecks need to be properly addressed, which include expensive high dimensional tensor contractions in the complex space 
(i.e. a mathematical space based upon complex numbers),
expensive inter-processor communication, and load imbalance. These issues can not simply be addressed by utilizing more processors for the calculations, but rather, on the other hand, relying on the developments of new computational strategy, new numerical solvers, and new computation infrastructures that integrate both the novel computer science techniques and highly scalable computing resource. Here, in this work, we present our effort to develop a numerical library that tackles these issues for achieving a highly scalable and efficient GFCC approach. We will first briefly review the GFCC theory/methodology, and give an overview of the entire workflow. After pointing out the specific bottlenecks in the workflow, 
%
we will discuss how we tackle these bottlenecks in the implementation details.
Also, we will present a profiling analysis of our implementation of the optimized GFCC approach for its parallel performance on large computing facility~\cite{olcfsummit}. 
Finally, to highlight the capability of our GFCC implementation, we  perform  for the first time the GFCCSD calculations employing over 800 basis functions for computing the many-body electronic structure of the fullerene C60 molecule covering up to $\sim$ 25 eV near-valence spectral region.

\section{Theory}

For a review of the GFCC method employed in this work, we refer the readers to Refs. \cite{nooijen92_55, nooijen93_15, nooijen95_1681,meissner93_67,kowalski14_094102, kowalski16_144101,kowalski16_062512, kowalski18_561,kowalski18_4335, kowalski18_214102}. 
Briefly, the matrix element of the analytical frequency dependent Green's function of an $N$-electron system at the frequency $\omega$ can be expressed as
\begin{widetext}
\begin{eqnarray}
G_{pq}(\omega) =
\langle \Psi | a_q^\dagger (\omega + ( H - E_0 ) - i \eta)^{-1} a_p | \Psi \rangle + 
\langle \Psi | a_p (\omega - ( H - E_0 ) + i \eta)^{-1} a_q^\dagger | \Psi \rangle
\label{gfxn0}
\end{eqnarray}
\end{widetext}
Here $H$ is the electronic Hamiltonian of the $N$-electron system, $| \Psi \rangle$ is the normalized ground-state wave function of the system, $E_0$ is the ground state energy, $\eta$ is the broadening factor introduced numerically to provide the width of the computed spectral bands, and the $a_p$ ($a_p^\dagger$) operator is the annihilation (creation) operator for electron in the $p$-th spin-orbital (we use $p,q,r,s,\ldots$ for the general spin-orbital indices, $i,j,k,l,\ldots$ for the occupied spin-orbital indices, and $a,b,c,d,\ldots$ for the virtual  spin-orbital indices). Integrating the Green's function formulation into the bi-orthogonal coupled cluster (CC) formalism, the generated GFCC formulation can then be expressed as
\begin{widetext}
\begin{eqnarray}
G_{pq}(\omega) = 
\langle\Phi|(1+\Lambda) \overline{a_q^{\dagger}} (\omega+\bar{H}_N - \text{i} \eta)^{-1} \overline{a_p} |\Phi\rangle +
\langle\Phi|(1+\Lambda) \overline{a_p} (\omega-\bar{H}_N + \text{i} \eta)^{-1} \overline{a_q^{\dagger}} |\Phi\rangle
\label{gfxn1}
\end{eqnarray}
\end{widetext}
with $|\Phi\rangle$ being the reference function, and the normal product form of similarity transformed Hamiltonian $\bar{H}_N$ being defined as $\bar{H} - E_0$. Here, the similarity transformed operators $\bar{A}$ ($A = H, a_p, a_q^{\dagger}$) are defined as $\bar{A} = e^{-T} A ~e^{T}$, and the cluster operator $T$ and the de-excitation operator $\Lambda$ are obtained from solving the conventional CC equations. By defining $\omega$-dependent many-body operators $X_p(\omega)$ and $Y_q(\omega)$
as
\begin{widetext}
\begin{eqnarray}
X_p(\omega) 
&=& X_{p,1}(\omega)+X_{p,2}(\omega) + \ldots = \sum_{i} x^i(\omega)_p  a_i  + \sum_{i<j,a} x^{ij}_a(\omega)_p a_a^{\dagger} a_j a_i +\ldots , \label{xp} \\
Y_q(\omega) 
&=& Y_{q,1}(\omega)+Y_{q,2}(\omega) + \ldots = \sum_{a} y^a(\omega)_q  a_a^\dagger  + \sum_{i,a<b} y^{i}_{a,b}(\omega)_q a_a^{\dagger} a_b^{\dagger} a_i +\ldots , \label{yq} 
\end{eqnarray}
\end{widetext}
(the dependence of $G$, $X$, and $Y$ on $\omega$ will not explicitly mentioned henceforth) to satisfy 
\begin{eqnarray}
(\omega+\bar{H}_N - \text{i} \eta )X_p|\Phi\rangle &=& \overline{a_p} |\Phi\rangle,  \label{eq:xplin} \\
(\omega-\bar{H}_N + \text{i} \eta )Y_q|\Phi\rangle &=& \overline{a_q^\dagger} |\Phi\rangle,  \label{eq:xplin} 
\end{eqnarray}
Eq. (\ref{gfxn1}) can be re-expressed by a compact expression
\begin{eqnarray}
G_{pq} = 
\langle\Phi|(1+\Lambda) \overline{a_q^{\dagger}} X_p |\Phi\rangle +
\langle\Phi|(1+\Lambda) \overline{a_p} Y_q |\Phi\rangle.
\label{gfxn2}
\end{eqnarray}
By truncating the many-body expansion of the cluster and mapping amplitudes (i.e. $T$, $\Lambda$, $X$, and $Y$) at the two-body level, the obtained approximate formulation of Eq. (\ref{gfxn2}) can be writen as
\begin{widetext}
\begin{eqnarray}
G_{pq} =  
\langle\Phi|(1+\Lambda_1+\Lambda_2) \overline{a_q^{\dagger}} (X_{p,1}+X_{p,2}) |\Phi\rangle +
\langle\Phi|(1+\Lambda_1+\Lambda_2) \overline{a_p} (Y_{q,1}+Y_{q,2}) |\Phi\rangle 
\label{gfxn3},
\end{eqnarray}
\end{widetext}
which is the so-called GFCCSD approximation (GFCC with singles and doubles) with $X_{p,1}$($Y_{q,1}$ or $\Lambda_1$) and $X_{p,2}$($Y_{q,2}$ or $\Lambda_2$) being one- and two-body component of $X_{p}$ ($Y_{q}$ or $\Lambda$) operator, respectively. The  electronic structure of the system is then captured by the computed spectral function from the GFCCSD approximation that is given by the trace of the imaginary part of the retarded GFCCSD matrix,
\begin{equation}
A = - \frac {1} {\pi} \text{Tr} \left[ \Im\left({\bf G}^{\text{R}} \right) \right] 
= - \frac {1} {\pi} \sum_{p} \Im\left(G_{pp}^{\text{R}} \right)~. \label{specfxn}
\end{equation}
Eqs. (\ref{gfxn0})$-$(\ref{specfxn}) describe the working equations to obtain the GFCC matrix element and spectral function at a specific frequency. For a broad frequency regime with a high frequency revolution, it would require the GFCC calculation at tens or hundreds of frequencies that constitutes a sizeable prefactor for the calculation.

\begin{figure}[htbp]
\centerline{\includegraphics[width=0.4\textwidth]{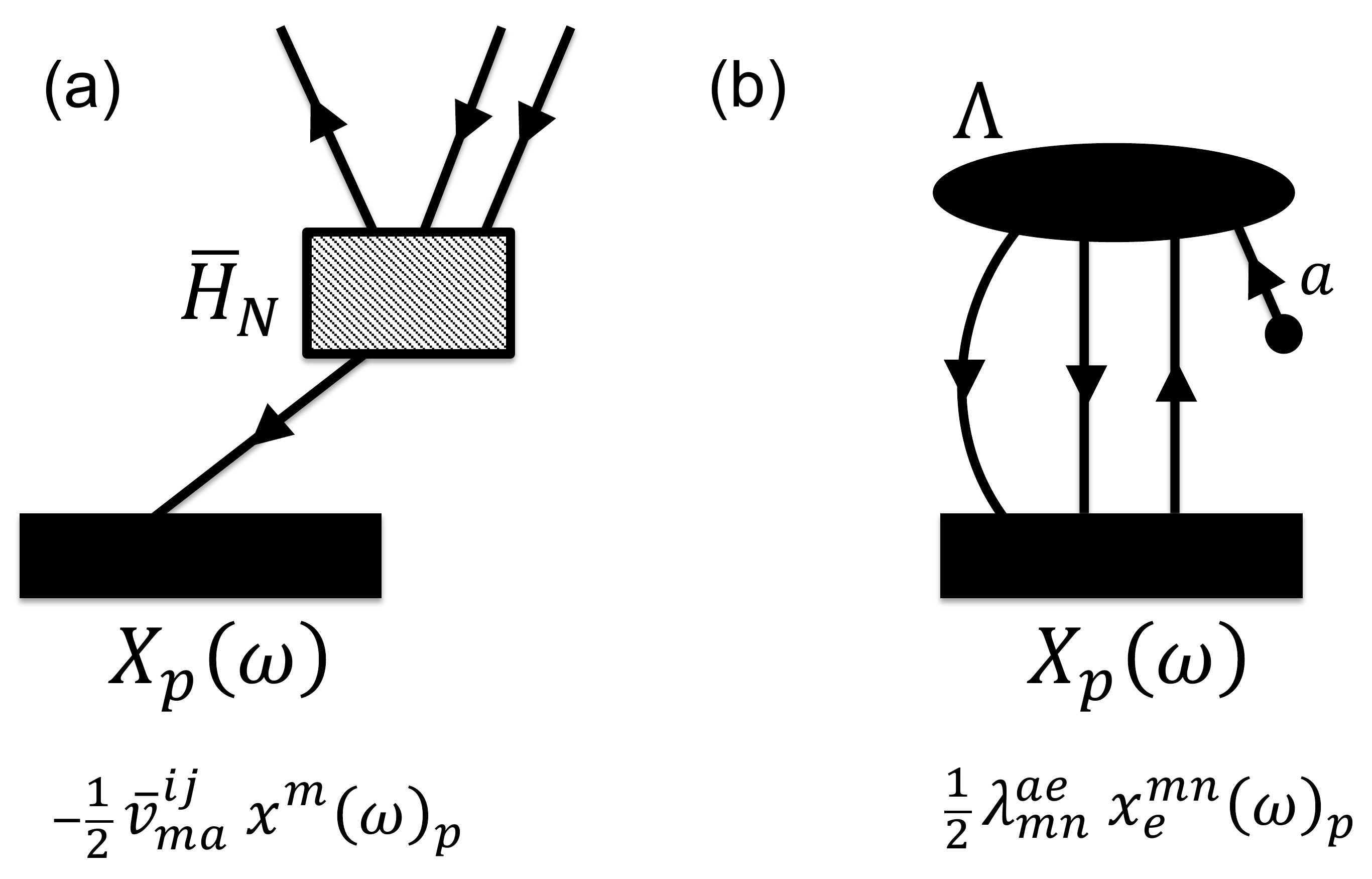}}
\caption{Examples of diagrams and corresponding tensor contractions contributiong to Eqs.(\ref{eq:xplin}) (inset (a))  and (\ref{gfxn2}) (inset (b)).}
\label{diagrams}
\end{figure}

The working equations for GFCC formulations can also be derived using various diagrammatic techniques. The corresponding  diagrams (typical examples of diagrams defining equations for $X_p$ operator and matrix elements of CC Green’s function are shown in Fig.\ref{diagrams}) can be translated into the form of tensor contractions involving tensors describing interactions included in the Hamiltonian operator, ground-state CC amplitudes, and amplitudes defining $X_p$ operators. Efficient implementations of these complicated expressions require specialized form of tensor libraries to provide tools  for distributing multidimensional tensors across the network, performing tensor contraction in parallel, and  utilizing computational resources offered by existing GPU architectures.

In order to reduce the formal computational cost of the GFCC method over a broad frequency regime, further approximation has been made in our calculation through the model-order-reduction (MOR) technique\cite{peng19_3185}. Specifically, one can construct an effective Hamiltonian based on the orthonormal subspace $\mathbf{S}=\{\mathbf{v}_1, \mathbf{v}_2, \ldots, \mathbf{v}_m\}$ (with its dimension $m$ being much smaller than the dimension of the original Hamiltonian) such that the conventional GFCC linear system over a broad frequency regime can be approximated through a more easily solvable model linear system over the same frequency regime,
\begin{widetext}
\begin{eqnarray}
\left\{ \begin{array}{ccl}
(\omega - \text{i} \eta + \hat{\bar{\textbf{H}}}_N) \hat{\textbf{X}} & = & \hat{\textbf{b}}, \\
\hat{\mathbf{G}}^{\text{R}} & = & \hat{\mathbf{c}}^{\text{T}} \hat{\textbf{X}},
\end{array}\right. ~~~\text{and}~~~
\left\{ \begin{array}{ccl}
(\omega + \text{i} \eta - \hat{\bar{\textbf{H}}}_N) \hat{\textbf{Y}} & = & \hat{\textbf{b}}, \\
\hat{\mathbf{G}}^{\text{A}} & = & \hat{\mathbf{c}}^{\text{T}} \hat{\textbf{Y}},
\end{array}\right.
\label{model}
\end{eqnarray}
\end{widetext}
Here $\hat{\bar{\textbf{H}}}_N = \textbf{S}^{\text{T}} \bar{\textbf{H}}_N \textbf{S}$, $\hat{\textbf{X}} = \textbf{S}^{\text{T}} \textbf{X}$, $\hat{\textbf{Y}} = \textbf{S}^{\text{T}} \textbf{Y}$, $\hat{\textbf{b}} = \textbf{S}^{\text{T}} \textbf{b}$, and $\hat{\mathbf{c}}^{\text{T}} = \mathbf{c}^{\text{T}} \textbf{S}$ with the columns of $\textbf{b}$ corresponding to $\overline{a_p} |\Phi\rangle$ or $\overline{a_q^\dagger} |\Phi\rangle$, and the columns of $\textbf{c}$ corresponding to $\langle \Phi | (1+\Lambda) \overline{a^\dagger_q}$ or $\langle \Phi | (1+\Lambda) \overline{a_p}$, respectively. 

\section{GFCC Workflow and Bottlenecks}
\begin{figure}[htbp]
\centerline{\includegraphics[width=0.6\textwidth]{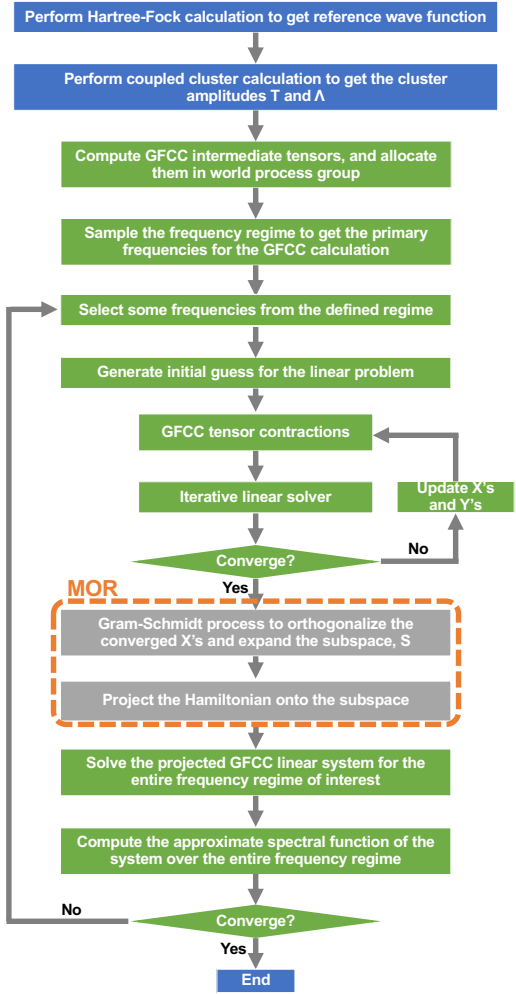}}
\caption{The GFCC workflow for many-body electronic structure calculation of a quantum systems.}
\label{workflow}
\end{figure}

A typical GFCC workflow for calculating the many-body electronic structure of a quantum system is shown in Fig. \ref{workflow}. As shown, prior to the practical calculation of GFCC matrix, the conventional Hartree-Fock (HF) and coupled cluster (CC) calculations need to be performed to get converged reference wave function and cluster amplitudes $T$ and $\Lambda$. The computational cost of the conventional HF calculation scales as $\mathcal{O}(N^3)$, while the computational cost of the conventional CC calculation depends on the truncation level in the many-body expansion of the cluster amplitudes. For coupled cluster singles and doubles (CCSD) calculation, where only one- and two-body terms are included in $T$ and $\Lambda$, the computational cost scales as $\mathcal{O}(O^2V^4)$ ($O$ denotes the number of the occupied orbitals, $V$ denotes the number of virtual orbitals, and $O+V=N$).
To construct the retarded GFCC matrix (similar for the advanced GFCC matrix), the key step is to use iterative linear solver to solve Eq. (\ref{eq:xplin}) for $X_p$ (the inner cycle of Fig. \ref{workflow}). For given orbital $p$ and frequency $\omega$, the computational cost for each iteration in the GFCCSD level approximately scales as $\mathcal{O}(O^3V^2)$. 
The MOR step is performed after solving the linear equations (as marked by the red dashed frame in Fig. \ref{workflow}), which involves two steps, (i) a Gram-Schmidt (GS) orthogonalization for newly obtained $X_p$'s with respect to the previous orthonormal vectors (if any) to generate/expand an orthonormal subspace, and (ii) a projection of the original similarity transformed Hamiltonian to the aforementioned subspace. The cost of GS step scales as $\mathcal{O}(m^2O^2V)$, while the cost of the projection scales as $\mathcal{O}(mO^3V^2)$. 
Finally, after performing MOR, one then needs to solve the projected GFCC linear system and compute the spectral function, of which the cost approximately scales as $\mathcal{O}(N_{\omega}m^3)$ with $N_{\omega}$ being the total number of frequencies. 
It is worth mentioning that, as discussed in Ref. \citenum{peng19_3185}, the rank of the subspaces (i.e. the dimension $m$) depends on the required accuracy, the interpolated frequency regime, and frequency interval. Typically, for a designated frequency regime being interpolated, the number of levels in the GFCC loop is usually less than five, and the rank of the subspace is $(5\sim33)\times N$ where $N$ is the number of MOs.

From the above workflow and scaling analysis, the bottleneck of performing a GFCC calculation mainly comes from three parts, (i) CC calculations to get the cluster amplitudes, (ii) GFCC iterations, and (iii) the projection of GFCC linear system to the constructed orthonormal subspace. Based on our experience, both CC and GFCC calculations are bound to (a) intensive compute and communication and (b) performance of the iterative solver. Regarding (a), there are many similar tensors and tensor contractions between CC and GFCC calculations, which indicate a universal tensor algebra library may be applied. Our effort of designing and applying such a tensor algebra library specific for the many-body methods will be detailed in the following section. For (b), it is worth mentioning that the fundamental difference from the mathematical point of view between the CC equations and the GFCC equations. The CC equations used to determined the cluster amplitude is a set of coupled energy-independent non-linear algebraic equations, while the GFCC equation is linear and $\omega$-dependent. The difficulty of efficiently solving these linear and nonlinear equations in the GFCC workflow comes from the fact that there is no unique solver that is able to efficiently deal with both linear and non-linear systems, and multiple solvers need to be included in the workflow to avoid the severe load imbalance (caused by failures to converge in some scenarios using certain solvers), and to reach a balance between efficiency and stability. 

According to previous studies\cite{pulay80_393,pulay82_556}, especially the early self-consistent field and ground state coupled cluster studies, the direct inverse of iterative subspace (DIIS) solver usually exhibits faster or even super-linear convergence performance for non-linear equations. In our pilot GFCC calculations, we have also applied DIIS to solve the GFCC linear equations for small molecular systems such as CO and N$_2$, and found that the DIIS solver can be faster than the Lanczos iterative method in computing the valence spectral function. The difficulty of biconjugate gradient (BiCG) method (built on the unsymmetric Lanczos iterative methods) of converging to the solution of the $\omega$-dependent GFCC linear equation has also been reported in the spectral studies of single atoms \cite{matsushita18_034106}. Indeed, in the BiCG method, the residual does not reduce monotonically, and the convergence of the linear equation is not guaranteed. However, this doesn't mean that DIIS is superior to the Lanczos iterative method, and the difference often comes with how the initial guess and preconditioner are applied for different solvers. From our own experience, the difficulty of applying DIIS for solving the $\omega$-dependent GFCC linear equations often emerges when computing the shake-up states in the core regime where the higher order elements in the $X_p$ amplitude becomes significant but the quality of the initial guess generated from low order perturbation is poor and the conventional preconditioner (inverse of the diagonal of the Hamiltonian) is close to singular. Alternatively, one can choose non-preconditioning. A popular choice is the generalized minimal residual (GMRes) method \cite{saad86_856}, where the original linear equation is projected to the orthonormal Krylov subspace generated from power iterations to get approximate solution with minimal residual. Employing GMRes, the residual decreases monotonically, and in principle, the GMRes will converge to the exact solution after at most $\mathcal{O}(O^2V)$ steps. According to our observation, the GMRes for solving the GFCC linear equations is quite stable. Even though the GMRes might be slower than other solvers in some scenarios, we haven't encountered any situation so far where GMRes is unable to converge within a reasonable number of iterations, in particular when converging some shake-up states in the core regime. Unlike DIIS whose cost is constant about $\mathcal{O}(O^2V)$, the cost of GMRes grows as $\mathcal{O}(n^2O^2V)$ (with $n$ being the iteration number). In comparison with other linear solvers, GMRes is the only Krylov method that works for general matrices (note that the similarity transformed hamiltonian in GFCC is non-symmetric).

\section{Tensor Algebra for Many-body Methods}

\begin{figure}[htbp]
\centerline{\includegraphics[width=0.45\textwidth]{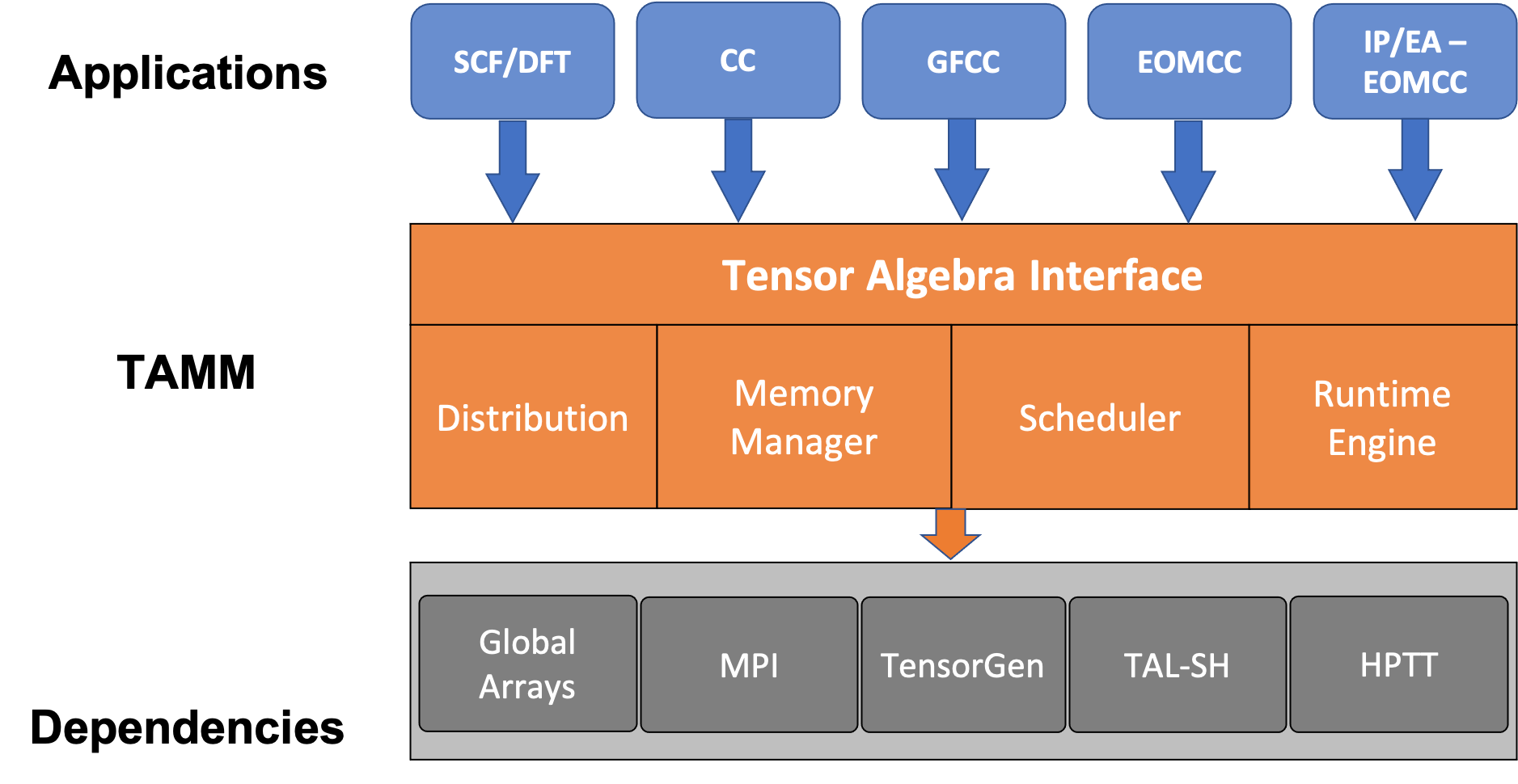}}
\caption{The TAMM architecture.}
\label{TAMMworkflow}
\end{figure}

As aforementioned, the most expensive parts of the GFCC approach are associated with the high dimensional tensor contractions. For example, the most expensive tensor contraction in the GFCCSD approach can be expressed as 
\begin{equation}
A(i,j,k,l) += B(i,j,m,n)\times C(m,n,k,l) \;,
\label{cont1}
\end{equation}
in which Einstein summation convention over the repeated indices is invoked, and the involved tensors ($A$, $B$, and $C$) have four dimensions with the length of each dimension being $\sim\mathcal{O}(N)$. These multi-dimensional tensor contractions are both compute and communication intensive. Previously, to ease these demands for large many-body calculations and new many-body theory developments, several specialized parallel tensor algebra systems have been developed
\cite{hirata039887, hirata2006symbolic, deumens2011software, deumens2011super, solomonik2013cyclops, solomonik2014massively, calvin2015scalable, peng2019coupled}.
Progress has been made for automated code generators, memory reduction, and the real-space many-body calculations. To further manipulate the tensor contractions on the complex space as required by the new GFCC approach, a more efficient tensor algebra library designed for generalized many-body calculations is needed.

The Tensor Algebra for Many-body Methods~(TAMM) library
\cite{mutlu2019toward}
provides such an infrastructure 
(see Fig. \ref{TAMMworkflow})
to achieve a scalable performance--portable implementation of key many-body methods on exascale supercomputing platforms.
Briefly speaking, the TAMM infrastructure is flexible in allowing the user to specify and manipulate tensor distribution, memory management, and scheduling of tensor operations, and supporting full complex and complex-real mixed operations on tensors that is mainly driven by the GFCC developments (the imaginary broadening in Eq. \ref{gfxn1} makes the GFCC approach a many-body theory on the full complex space being different from other excited state coupled cluster approaches). The TAMM infrastructure is implemented using Global Arrays (GA)
\cite{ga1994,ga1995,ga1996}
and MPI for scalable parallelization on distributed memory platforms and using  optimized libraries for efficient intra-node execution of tensor operation kernels on CPUs and accelerators.

TAMM takes high-level expressions describing computations on block-sparse tensors,
decomposes these into a set of dependent operations that are then passed to a backend for scheduling and execution. High performance is obtained in the backend by focusing upon a small number (several tens) of kernels that are extensively optimized by the vendor libraries, or by code generation plus auto-tuning, or by hand tuning. The TAMM API includes these high-level expressions and that of the distributed, block-sparse tensor.

Execution of operations using the TAMM library involves multi-granular dependence
analysis and task-based execution. 
To leverage parallelism across a set of tensor operations, a dependency analysis scheme (called the levelizer) in the TAMM scheduler splits the tensor operations into independent execution levels. By analyzing data dependences between a group of tensor operations, the tensor operations are organized into levels. Operations in the same level can be executed concurrently with synchronizations between levels. This allows for multiple operations to be executed at the same time, exposing more parallelism, and improving processor utilization.
The operations that can be scheduled in parallel are executed in a single program multiple data (SPMD) fashion. The execution is compatible with MPI and the operations are collectively executed on a given MPI communicator.

Each operation is further partitioned into tasks. The tasks that constitute an operation are produced using task iterators. Each task computes a portion of the operation, typically a contribution to a block of data in the output tensor. Until it begins execution, a task is migratable and can be scheduled for execution on any compute node or processor core. Once execution of a task begins, the data required by the task are transferred to its location. At this point, the task is bound to the process in which it is executing and cannot be migrated. 

TAMM uses a GPU execution scheme (with TALSH \cite{TALSH} as the underlying GPU tensor algebra engine) where we make use of localized summation loops to limit the transfer of output blocks from GPUs to CPUs. By keeping the output block that is being updated by multiple input tensor blocks on the GPU until all updates are finished, we are able to reduce the data transfer between CPUs and GPUs.  TAMM also uses a non-blocking GPU kernel launch scheme to enable data transfer/compute overlap between successive summation loop iterations.

The productivity and performance benefits of TAMM are presented in upcoming evaluation section.

\section{Implementation Details}
In our current implementation of the GFCC approach, to ease the memory/storage demands and increase the data locality (to reduce communication), a well-controlled on-the-fly pivoting Cholesky decomposition (CD) is performed for the four dimensional atomic-orbital based electron repulsion integral (ERI) tensors right after the relatively cheap Hartree-Fock calculations. According to the previous study \cite{kowalski17_4179}, by using generated three-index Cholesky vectors instead of four-index ERI tensors one can bypass the $\mathcal{O}(N^5)$ tensor transformation from atomic-orbital to molecular-orbital space with two $\mathcal{O}(N^4)$ steps, and significantly reduce the storage requirement from $\mathcal{O}(V^4)$ for the four-index ERIs to $\mathcal{O}(V^2N)$ for Cholesky vectors in the ground- and excited-state CC calculations. 

Before entering the main GFCC loop, the major contractions have been designed to maximize the number of constant intermediate tensors of small- and medium-size (for GFCCSD method, intermediate tensor size needs to be $\le\mathcal{O}(O^2V^2)$). These intermediate tensors will be pre-computed to ease the operation demands in the main GFCC loop. 

Our main GFCC loop is shown in Algorithm~\ref{alg:gfcc_pseudo} where we are focused on the GFCCSD method. At the high level, there is an outer loop that checks for the convergence of the entire GFCCSD calculation. We refer to iterations of this loop as $levels$. Each $level$ consists of two loops: one that goes over the frequencies ($\omega$'s) for a given $level$ and the second loop goes over all the orbitals ($p$'s). 
In the first loop, $\omega$'s are sampled following the adaptive midpoint refinement strategy described in Ref. \cite{vanbeeumen17_4950}. 
In the second loop, the GFCCSD singles and doubles equations need to be solved for all ($\omega$,$p$) pairs in a given level. Here we are focused on Lines 4$-$15 of Algorithm ~\ref{alg:gfcc_pseudo} which constitute the most expensive tensor contraction and communication, as well as the most severe load imbalance, of each $level$ in the  overall iterative GFCCSD calculation.

The upper bound on the maximum number of $level$s ($L_{max}$) in a GFCC calculation is determined as follows. Given a frequency regime $\left[\omega_\text{min}, \omega_\text{max}\right]$, and a desired frequency resolution $\Delta\omega$, The total number of available frequencies, $N_{\text{tot},\omega}$, in the regime is $N_{\text{tot},\omega} = \frac{\omega_\text{max}-\omega_\text{min}}{\Delta\omega}+1$, and the maximum number of frequencies at $level$ $L$, $N_{\text{max},\omega}(L)$, is given by 
\begin{equation}
    N_{\text{max},\omega}(L) = \left\{ \begin{array}{lcl}
    3   &  \mbox{for} & L = 1,\\
    2^{L-1} &  \mbox{for} & L > 1.
    \end{array}\right.
\end{equation}
Since $\sum_{L=1}^{L_\text{max}} N_{\text{max},\omega}(L) \le N_{\text{tot},\omega}$, the upper bound of $L_\text{max}$ is
\begin{equation}
    L_\text{max} \le \log_2 \left( 
    \frac{\omega_\text{max}-\omega_\text{min}}{\Delta\omega}
    \right). \label{Lmax}
\end{equation}
According to Eq. (\ref{Lmax}), take a frequency regime of [-0.8,-0.4] a.u. for an example, if $\Delta\omega$=0.01 a.u., then $L_\text{max}\le 5$.

\begin{algorithm*}[htbp]
	\small
	\SetKw{To} {\textbf{to}}
	\SetKw{Step} {\textbf{step}}
	\SetKwInOut{Input} {input}
	\SetKwInOut{Output} {output}
	\SetKwRepeat{Do}{do}{while}%
	int $level$ = 0; \\
	Setup process groups based on resources provided to the application run \\
	\While{not converged} {
	    \For {$\omega$ = 0 \To $n\_freq$} {
	        task\_list = ($\omega$,p) for  p = 0 \To $n\_orbitals$  \\
	        divide all tasks in task\_list across process groups \\
	        \For {p = 0 \To $n\_orbitals$} {
	           each process group executes a task ($\omega$,p) \\
	           determine task ($\omega$,p) for a process group ${\rm PG}_i$ \\
                setup intermediate and output tensors \\
                compute initial guess \\     
                \While{not converged} {
                    solve GFCC singles equations $\mathcal{O}(O^3V)$ \\
                    solve GFCC doubles equations $\mathcal{O}(O^3V^2)$ \\ 
                    GMRES micro loop $ngmres * \mathcal{O}(O^3V^2)$
                }
                write output tensors for ($\omega$,p) to disk \\     
            }
            synchronize across all process groups \\
	    }
	    Gram-Schmidt ($\mathcal{O}(m^2O^2V)$) \\ 
	    project GFCC linear systems onto the subspace constructed for all $(\omega,p)$'s in current $level$ ($\mathcal{O}(O^3V^2)$) \\
	    compute spectral function ($\mathcal{O}(N_{\omega}m^3)$) \\
	    determine convergence (OR) list of frequencies to be processes in next $level$ \\
	    $level$++;
	}
	\caption{Retarded GFCCSD Approach}
	\label{alg:gfcc_pseudo}
\end{algorithm*}
We use the TAMM library to implement the GFCCSD approach shown in Algorithm~\ref{alg:gfcc_pseudo}.
The task list in a given $level$ is the list of all ($\omega$,$p$) pairs that need to be processed in that $level$.
Since the computation of all tasks in a given $level$ are independent, we divide all the $p's$ for a given $\omega$ across process groups that are set up at the beginning of the calculation. All process groups are of the same size. The size of a process group for computing each task is determined automatically for a given problem size and the resources provided for that run. The size of each process group can also be provided as an input parameter for a given GFCCSD calculation. We present a detailed analysis of using different sized process groups for each task in the upcoming evaluation section. 

It is worth mentioning that due to the difference in the orbitals, the converging performance for different ($\omega$,$p$) pairs would become different and lead to severe load imbalance. 
To see that, for a given $\omega$ the spectral function is the trace of the GFCC matrix (sum of the diagonal elements), therefore if the $\omega$ is located on the valence regime, due to the relatively weak coupling between the core orbitals and valence orbitals, the diagonal elements contribution from the core orbital will be negligible. In terms of the convergence performance, the corresponding GFCC calculations will usually converge within one iteration or two. Fast convergence can also be observed for computing valence contribution to the core spectral function for the same reason. However, to compute the contribution from the orbitals that are lying close to the $\omega$, the convergence will involve more iterations, in particular when computing the shake-up states where the double excitation dominates the ionized wave function. 

\begin{table}[]
    \centering
    \resizebox{\columnwidth}{!}{%
    \begin{tabular}{|c|c|c|c|c|}
    \hline
Nodes & Basis & BasisSet & NWChem & TAMM \\
\hline  
    100 & 6-31G & 424 & 2.8 & 0.22 \\
    220 & cc-pVDZ & 737 & 13 & 0.58 \\
    256 & aug-cc-pVDZ & 1243 & 74 & 2.5 \\ 
    \hline
    \end{tabular}
    \caption{CCSD performance compared to NWChem for the Ubiquitin DGRTL fragment on OLCF Summit. Time per CCSD iteration is given in minutes.
        NWChem has CPU only implementation. TAMM based Cholesky-CCSD uses GPUs. }
    \label{tab:ccsd}
    }
\end{table}

\begin{table}[]
\resizebox{\columnwidth}{!}{%
\begin{tabular}{|c|c|c|c|l|}
\hline
Impl. & \begin{tabular}[c]{@{}c@{}}NWChem\\ CCSD\end{tabular} & \begin{tabular}[c]{@{}c@{}}TAMM\\ Cholesky-CCSD\end{tabular} & \multicolumn{2}{c|}{TAMM-GFCCSD-IP}                          \\ \cline{4-5} 
                       &                                                                        &                                                                               & \multicolumn{1}{l|}{Closed-shell} & Open-shell                \\ \hline
SLOC                   & 11314                                                                  & 236                                                                           & 1700                             & \multicolumn{1}{c|}{2700} \\ \hline
\end{tabular}
\caption{SLOC (source lines of code) counts for different CCSD and GFCC implementations. For the TAMM-GFCCSD-IP implementations, the closed-shell case only deals with the alpha electrons in the system, while the open-shell case deals both alpha and beta electrons in the system. }
\label{tab:sloccount}
}
\end{table}

\begin{figure}
    \centering
    \includegraphics[clip,angle=0, width=0.45\textwidth]{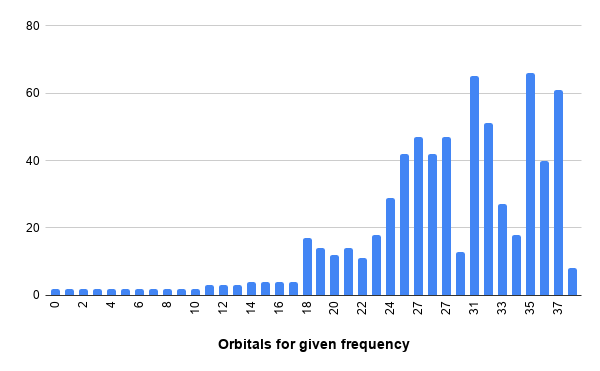}
    \caption{Number of iterations (y-axis) needed to converge  linear equations for $X_p$ operators corresponding to different orbitals of guanine for $\omega$=-0.4 a.u.}
    \label{guan}
\end{figure}

\begin{figure}
    \centering
    \includegraphics[clip,angle=0, width=0.45\textwidth]{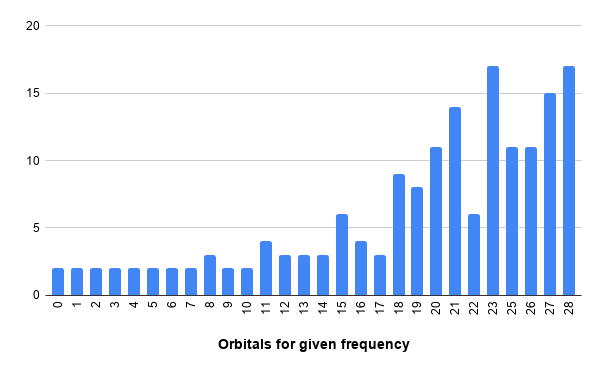}
    \caption{Number of iterations (y-axis) needed to converge linear equations for $X_p$ operators corresponding to  different orbitals of cytosine for $\omega$=-0.4 a.u.}
    \label{cyto}
\end{figure}

To further illustrate the load imbalance issue originated from the orbital difference on the contribution to the spectral function, we have exhibited the number of iterations needed to converge the GFCC linear equations using two small molecules, guanine and cytosine, at the valence frequency for all the occupied molecular orbitals in Figs. \ref{guan} and \ref{cyto}. As can be seen, for the probing frequency $\omega$ = -0.4 a.u., which is very close to the valence band edges of guanine and cytosine base molecules, there is a clear distinction between the core region and valence region in terms of the number of iterations. Due to the aforementioned weak coupling between the core molecular orbitals and valence molecular orbitals, the contribution of the core orbitals to the trace of the GFCCSD matrix is trivial, thus only one or two iterations are usually used to solve the GFCCSD linear equation, and the solving process is insensitive to the linear solver. Note that in Figs. \ref{guan} and \ref{cyto} there are also some valence orbitals possessing low number of iterations for solving the GFCCSD linear equations. This is because the spectral amplitude at the probing frequency ($\omega$=-0.4 a.u.) does not reach its prime. When the probing frequency is moving towards the ionization potentials that the molecular orbitals contribute the most to, the number of iterations will generally bounce back. For example, we have only used five iterations to converge the GFCCSD linear equation for molecular orbital \#38 of the guanine base at the probing frequency -0.4 a.u. If we change the probing frequency to -0.25 a.u. that is more close to the first IP value of guanine base, the number of iterations will go up to $\sim$40 iterations (the molecular orbital \#38 will contribute the most to the first IP of the guanine base).

To prevent any load imbalance between different process groups computing different orbitals, a process group that finishes the computation of a orbital earlier will be told to proceed to pick another orbital in the list of orbitals remaining to be computed for that frequency. It would not wait for other process groups to finish before proceeding to the next set of orbitals. However, there can still be a situation in which a few process groups are idle when the last set of orbitals for a given frequency are being processed as indicated by the barrier in Line 15 of Algorithm~\ref{alg:gfcc_pseudo}. In practice, we would not face this load imbalance for the last set of orbitals being processed for a given $\omega$. This is due to the fact that for a  GFCC calculation aimed at solving a real science problem, the number of orbitals are large enough such that it's enough to keep an entire modern supercomputer such as Summit busy just for computing a fraction of orbitals for a single frequency. We finish the entire GFCC calculation after a series of application restarts since there are no LCF machines that provide a job running time required to finish the GFCC calculation for even a single frequency in one attempt of the application run. The last set of orbitals might as well be processed in a separate job. 
After all orbitals for a given frequency are computed, the calculation proceeds to the next frequency in that $level$. Another optimization that is straightforward to implement in our current infrastructure is to process all orbitals for all the frequencies in a given $level$  simultaneously using  using process groups. 
In the current situation, the optimization brought by such an implementation will be greatly compromised due to a relatively short job walltime on LCF machines.

The current version of GFCCLib can be obtained at https://github.com/spec-org/gfcc/.

\section{Evaluation}

We evaluate the performance of the GFCC application on OLCF Summit~\cite{olcfsummit}.
Each Summit node has two 22-core POWER9 CPUs and 512GB of CPU memory. Each node is also equipped with 6 NVIDIA Volta GPUs with a total GPU memory of 96GB. The GFCC application was compiled using GCC 8.1 compiler, CUDA 10.1 toolkit, IBM Spectrum MPI 10.3, IBM ESSL 6.1 and BLIS 0.6.

Table~\ref{tab:ccsd} shows the performance of TAMM based Cholesky CCSD in comparison with the performance of the state-of-the-art CCSD in the open-source quantum chemical platform NWChem\cite{valiev2010nwchem}. The NWChem run was performed using all the CPU cores available on each node, while the TAMM-based implementation used 6 CPUs mapped to 6 GPUs on each node. As can be seen, the overall performance of TAMM-based Cholesky CCSD is up to $\sim$13 times faster than that of the NWChem run for the  closed-shell calculation of large DNA fragment~\cite{ubiquitin}.

The promising speed-up of TAMM-based Cholesky CCSD with respect to the NWChem CCSD module can be attributed to many factors including GPU acceleration, dependency analysis and scheduling of independent contractions in parallel to enable barrier-minimized scheduling across tensor operations, better data/tensor distribution enabled by Cholesky vectors, and locality-aware parallelization methodologies. 
Note that the Cholesky vectors are not used in the NWChem CCSD module (also there is no GFCC module in NWChem). 
Generally speaking, the utilization of Cholesky vectors can greatly reduce the memory/storage demands of the calculation, and therefore the communication demands \cite{kowalski17_4179}. In particular, the storage requirement associated with the two-electron integral tensors in the molecular orbital basis ($\mathcal{O}(N^4)$) can be avoided. Instead, only $\mathcal{O}(N^3)$ requirement is needed for storing the Cholesky vectors. Therefore, the $\mathcal{O}(N^5)$ atomic orbital (AO) to molecular orbital (MO) two-electron integral tensor transform will be bypassed, and replaced by the $\mathcal{O}(N^4)$ AO to MO Cholesky vector transformation.
It is worth mentioning that CD can help refactor and simplify a portion of the CCSD tensor contractions to yield lower computational scaling, but is unable to refactor the most expensive CCSD tensor contraction that scales as $\mathcal{O}(O^2V^4)$. Due to the large number of basis functions (and therefore high storage demand) and complexity of the conventional CCSD tensor contractions, we are currently unable to perform the conventional TAMM-CCSD calculation for C60 to single out the performance advantage only from utilizing the Cholesky vectors. On the other hand, the overall performance improvement associated with the utilization of Cholesky vectors in the CCSD approach depends on many factors such as the algorithm design, studied system, and decomposition threshold. In the previous CD-CCSD study of benzene trimer, utilizing Cholesky vectors (with 10$^{-4}$ threshold) can result in a modest speedup of 13\% \cite{sherrill13_2687}. 
To minimize load imbalance overheads, the CCSD execution is performed by grouping all operations in a single iteration, analyzing their dependencies, and grouping them into levels of independent operations. This scheduling of operations applies to all methods implemented using TAMM. Our GFCCSD implementation also benefits from all the optimizations mentioned. 

Here, we want to emphasize that, even though the NWChem calculation may also benefit from the GPU acceleration, the promising speedup were only limited to noniterative higher order many-body methods (note that the GPU implementation in the official releases of NWChem is only available for the perturbative part of the CCSD(T) formalism \cite{kowalski11_1316}; see Ref. \cite{kowalski13_1949} for a more recent example), and it is actually infeasible to make NWChem CCSD benefit from GPUs due to the large volume of code ($\sim$11,000 lines, see Table~\ref{tab:sloccount}) that need to be rewritten. On the other hand, since TAMM provides a high-level abstraction to compose sequence of tensor operations, the methods implemented using TAMM can simply specify the hardware resource that the operations need to be executed on - this allows the application code to be at a higher level while the implementation details of how the operations are executed are handled internally by TAMM. This also allows the application code to be ported to newer architectures when TAMM supports them without having to rewrite the application code itself.

\begin{table}[]
\centering
\resizebox{\columnwidth}{!}{%
\begin{tabular}{|c|c|c|c|c|c|c|c|}
\hline
\multicolumn{1}{|c|}{}  & \multicolumn{7}{c|}{\#Parallel Tasks}                                                                                                                                              \\ \hline
\multicolumn{1}{|c|}{}                     & \multicolumn{1}{c|}{1} & \multicolumn{1}{c|}{2} & \multicolumn{1}{c|}{3} & \multicolumn{1}{c|}{4} & \multicolumn{1}{c|}{10} & \multicolumn{1}{c|}{20} & \multicolumn{1}{c|}{40} \\ \hline
\multicolumn{1}{|c|}{Nodes, GPUs per task} & \multicolumn{7}{c|}{Time per iter per task}                                                                                                                                     \\ \hline
20,120                                    & X                      & X                      & X                      & X                      & 485                     & 493                     & 522                     \\
40,240                                     & X                      & X                      & 252                    & 250                    & 257                     & 264                     & 358                     \\
60,360                                     & X                      & 176                    & 176                    & 178                    & 180                     & 190                     & X                       \\
75,450                                     & 153                      & 147                    & 150                    & 148                    & 154                     & 163                     & X                       \\
100,600                                    & 119                      & 120                    & 120                    & 119                    & 126                     & 146                     & X                       \\
150,900                                    & 91                     & 92                     & 95                     & 99                     & 105                     & X                       & X                       \\
200,1200                                   & 77                     & 80                     & 81                     & 85                     & 88                      & X                       & X                       \\
250,1500                                   & 70                     & 68                      & 69                      & 71                     & X                       & X                       & X                     \\
\hline
\end{tabular}
\caption{The operation time per task (in seconds) for one GFCCSD iteration with different settings of number of nodes, number of GPUs, and number of tasks (orbitals) processed in parallel on OLCF Summit. The test system is C60 molecule with aug-cc-pVDZ basis set with a total of 1380 basis functions and 130 linear dependencies.}
\label{gfcc_scaling_data}
}
\end{table}

\begin{table}[]
    \centering
    \resizebox{\columnwidth}{!}{%
    \begin{tabular}{|c|c|c|c|c|c|}
    \hline
        Nodes, GPUs & get & compute  & add   & misc. & Total \\
        \hline
         75, 450 & 4.4 & 120 &  0.35 & 28.3 & 153 \\
         100, 600 & 2.5 & 89 &  0.2 & 27.3 & 119 \\
         150, 900 & 3 & 60 & 0.3  & 31.7 & 95 \\
         200, 1200 & 2.3 & 44.6 & 0.2  & 33 & 80 \\
         250, 1500 & 2 & 36 & 0.13  & 30 & 68 \\
         \hline
    \end{tabular}
    \caption{Performance analysis of a single iteration for task ($\omega$,p) = (-0.4,160). The table shows the time (in seconds) spent in getting input tensors, computing, and putting (or adding) to output tensors. Total time per iteration and time not spent in communication or computation (``misc.'' time, dominated by load imbalance at barrier) are also shown.  We observe that all tasks exhibit nearly identical runtime behavior.}
    \label{tab:taskprofile}
    }
\end{table}

To study how different number of tasks will affect the operation time of the GFCCSD calculation, we have performed a series of tests with different numbers of nodes (and GPUs) to record the time for one GFCCSD iteration, and the results are shown in Table \ref{gfcc_scaling_data}. We choose C60 molecules (with $C$1 symmetry) as our testing system. Employing the aug-cc-pVDZ basis set, the GFCCSD tests were performed with over 1,000 basis functions on OLCF Summit. Here, a guideline is provided on choosing the number of nodes and tasks for a large GFCCSD calculation.
In Table~\ref{gfcc_scaling_data}, each row corresponds to a particular configuration. For example, row 1 shows the performance when using 20 nodes per task for various number of parallel tasks. For the configuration using 20 nodes per task and 40 tasks (last entry in row 1), the calculation was run on 800 nodes (20x40) where 40 process groups were executing the 40 different tasks for a given frequency simultaneously. 

This table also provides insights into the weak and strong scaling behavior of the implementation. 
It is worth mentioning that weak and strong scaling behaviors enable studying the performance of a parallel application on supercomputers. Strong scaling\cite{strongscaling} allows us to measure the parallel efficiency of an application using the speedup computed by increasing the number of computing resources for a fixed problem size. Weak scaling\cite{weakscaling} on the other hand describes application behavior when both the number of computing resources and problem size are increased.
Going down a column of Table~\ref{gfcc_scaling_data}, the same number of tasks is  performed on an increasing number of total nodes, effectively strong scaling the computation. Along a row, more tasks are performed in parallel, with each task performed on the same number of nodes.  
This corresponds to weak scaling the computation. In general, GFCC requires good weak scaling behavior to support multiple GFCCSD calculations to be efficiently performed in parallel. However, improved strong scaling can enable the use of fewer parallel tasks, reducing the impact of load imbalance. We observe that weak scaling leads to a small increase in total time. For example, per-iteration time increases from 77 seconds to 88 seconds as we go from executing 1 task to 10 tasks in parallel, with each task running on 200 compute nodes. Importantly, while not perfectly scaling, the implementation scales both under the weak scaling and strong scaling regimes. Combining both scaling modalities allows the GFCCSD implementation to run on very large node counts without significant performance degradation.

To get further insights into the implementation's strong-scaling behavior, we profiled the execution of a single iteration of GFCCSD for varied node counts. Table~\ref{tab:taskprofile} shows the time spent in getting data from input tensors (labeled "get"), performing the local computation (``compute,'' much of it spent performing tensor contractions), and putting or adding the result to the output tensors (labeled ``add''),  and the total per-iteration time. Time not accounted in data communication or computation is shown under ``misc.'', which is dominated by load imbalances at each iteration. We observe that compute time improves nearly linearly with node count. However, communication time, though initially small, starts to form an increasingly large fraction of the total time. Significantly, miscellaneous time within each iteration begins to dominate the total work performed. In general, as the per-iteration time approaches a minute, non-compute times become significant. By combining weak and strong scaling, we mitigate this effect at large node counts.

From the table, we observe that at least 20 nodes are needed for such a large GFCCSD calculation, and given a fixed number of nodes and GPUs per task, the time per iteration time per task remains consistent when executing up to 20 orbitals in parallel, which thus indicates a good scalability. Beyond 20 parallel tasks, we see a significant performance hit in the time per iteration per task. 
This is due to the fact that the MPI processes across all process groups are communicating to fetch from the same input (cluster amplitudes) and intermediate tensors leading to network congestion. 
Though the tensors involved in the computation of a given task are created and destroyed within the process group computing the task, all process groups share the input and some intermediate tensors that were created on the world process group. One way to address this problem is to replicate the input and intermediate tensors where possible. If the process group size used to compute each task is large enough for a given system size, we can identify and replicate the global cluster amplitude tensors and GFCCSD intermediate tensors that are frequently accessed by all process groups. Note that these tensors usually have small or medium size ($<\mathcal{O}(O^2V^2)$). The implementation of this replication scheme is still under intensive development, and will be further discussed in our future work.

\begin{figure}[htbp]
\centerline{\includegraphics[width=0.45\textwidth]{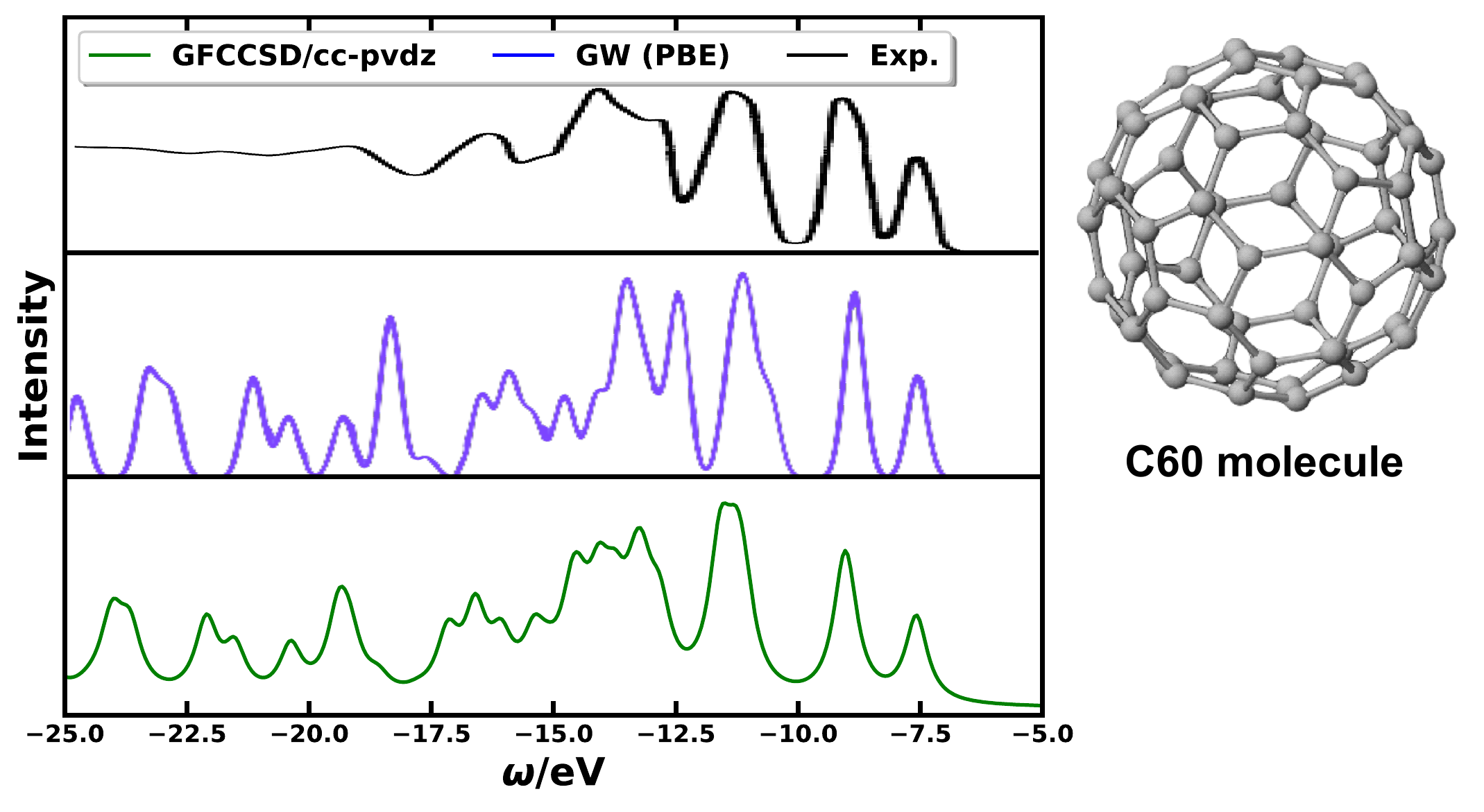}}
\caption{The computed spectral function of C60 molecule at the GFCCSD level, and in comparison with single-particle $GW_0$ (PBE) and experimental spectrum. The GW spectrum were adopted from Ref. \citenum{qian15_245105}. The experimental photoelectron spectrum of C60 were adopted from Ref. \citenum{benning92_6899}. In the GFCCSD calculation of C60 molecule, the total number of frequency points used to construct the subspace is nine, and the dimension of the subspace $m$ is $\sim10,800$.}
\label{specfxn}
\end{figure}

\section{Testing on C60 complex molecule}

To test the capability of the developed GFCC library, we choose the fullerene C60 molecule and compute its spectral function in a broad near-valence regime for the first time with the many-body GFCCSD method.
Since we are focused on the post-Hartree-Fock calculation in the valence region, Dunning's correlation-consistent polarized valence-only double-zeta, cc-pVDZ, basis set \cite{dunning89_1007} has been used in the calculation. Note that larger Dunning basis sets can be further used for converging GFCCSD calculations systematically to the complete basis set limit.

Our computed spectrum is shown in Fig. \ref{specfxn} where we also compared our results with the $GW_0$ (PBE) and experimental results. The lowest ionization potential (IP) computed from GFCCSD/cc-pVDZ is about -7.58 eV which is almost identical to the value of -7.6$\pm$0.2 eV from the experiment\cite{benning92_6899}, and the deviation is within the chemical accuracy ($\sim$1 kcal/mol, or 0.043 eV). The GFCCSD many-body results is superior to the single particle $GW_0$ results. The latter depends on the density functionals used in the calculation\cite{qian15_245105}. Using local-density approximation (LDA), the reported IP values range from -7.28 eV to -8.22 eV, and using generalized gradient approximation (GGA,\cite{langreth83_1809,becke88_3098,perdew92_6671} e.g. the Perdew--Burke--Ernzerhof (PBE) functional\cite{PBE96_3865}), the IP value was reported to be -7.37 eV. For higher IPs, in particular in the [-17.5,-5.0] eV regime, the $GW_0$ single particle results red-shift the entire spectrum by $\sim$ 0.5 eV with respect to the experimental spectrum, while the GFCCSD spectrum shows excellent agreement with the experiment. It is worth mentioning that the GFCCSD results can be further improved by including higher order amplitudes in the GFCC approach, and/or using larger basis set. Regarding the higher order GFCC approach (e.g GFCC with singles, doubles, and triples, GFCCSDT), sometimes only only a portion of higher order terms can greatly improve the description of the many-body states\cite{kowalski18_214102}. Regarding the basis set effect, it seems employing more diffuse basis set will not significantly change the IPs of C60. According to the previous EOM-CC study (same theoretical level as GFCC)\cite{kowalski14_074304}, the lowest IP of C60 will only be red-shifted by $\sim$0.08 eV if we switch from cc-pVDZ basis to aug-cc-pVDZ basis. 

By and large, GFCCSD approach (shown in Fig. \ref{specfxn}) has produced an accurate description of the electronic structure of the C60 molecule spreading over ~20 eV in the near valence energy regime. Accurate and efficient many-body GFCC routine calculations for other fullerene molecules and their derivatives can be anticipated in the near future. The accurate description of the electronic structure (in terms of peak positions and amplitudes) of these molecular systems will not only benefit a clear explanation of the static structure-property correlation, but also provide a better understanding of the electron dynamics (through the Fermi's golden rule) explaining the vital charge and energy flow when applying these systems in the nanostructured devices.

\section{Comparison with Prior Work}
The TAMM implementation  of the GFCCSD formalism discussed here is the only reported implementation of the GFCC formalism capable of carrying out calculations for hundreds of correlated electrons with $>$1,000 basis functions. As such, it provided  a unique capability that can be used not only in the simulations of spectral functions but also as an integral component of complex quantum embedding workflows. To answer the scientific questions of (i) how the ionizations of DNA fragments and their features change as the system expands and (ii) how the many-body coupled cluster description would be different from the single-particle picture and lead us to a more generalized near-valence ionization picture of longer DNA sequence, we expect the TAMM GFCCSD implementation to be further applied to the simulations of spectral function for very large DNA fragments in the complex environment with the aid of quantum embedding scheme. The anticipated effort will go beyond the thus-far largest GFCCSD simulation (as we mentioned in the introduction, the latter was performed for a gas phase DNA hexamer without the perturbation from the solution environment  \cite{peng20_011101}).

Compared to prior implementations of correlated Green’s function formulations such as various GW and ADC(n) approximations, the present GFCC implementation provides significant advance to (i) simulate larger and more complex quantum systems, and (ii) understand the ionization process that can support various photoelectron spectroscopy studies.  Additionally, a unique footprint of the present implementation is its flexibility in employing various representations of two-electron integrals (such as Cholesky Decomposition) and parallel models.

Furthermore, the present GFCC implementation pave the way for real applications of GFCC approach to real systems and to arbitrary frequency regimes. Also, controlling the accuracy of the Dyson equation (Green’s function and self-energy matrices are complementary quantities in the sense of this equation) on the CC level is much easier at the level of Green’s function than at the level of the self-energy operator. Recent studies\cite{hirata17_044108} of the perturbative self-energy expansions also demonstrated that these expansions may be characterized in some situations by a slow convergence rate. Additionally, the MOR methods adapted for the GFCC approach were validated only for small-dimensionality problems in the previous applications.\cite{peng19_3185} In the application of present GFCC implementation, these techniques were successfully extended  to high-dimensionality problems allowing for successful identification of not only main peaks (which can be identified by lower-order methods such as GW formalisms) but also satellite states that are much harder to capture in lower-order methods. To our best knowledge, there are no existing  GFCC implementations that enable the routine calculations with $>$1,000 basis functions for arbitrary frequency regimes. From the technical viewpoint, the novel interdisciplinary engineering of tensor contractions library (TAMM), MOR algorithm, efficient compression algorithms for two-electron integrals, and GPUs into a many-body theoretical framework enables us even with the existing implementation to perform simulations for even bigger systems described by 2,000$-$3,000 orbitals.


\begin{figure}[tp]
\centerline{\includegraphics[width=0.45\textwidth]{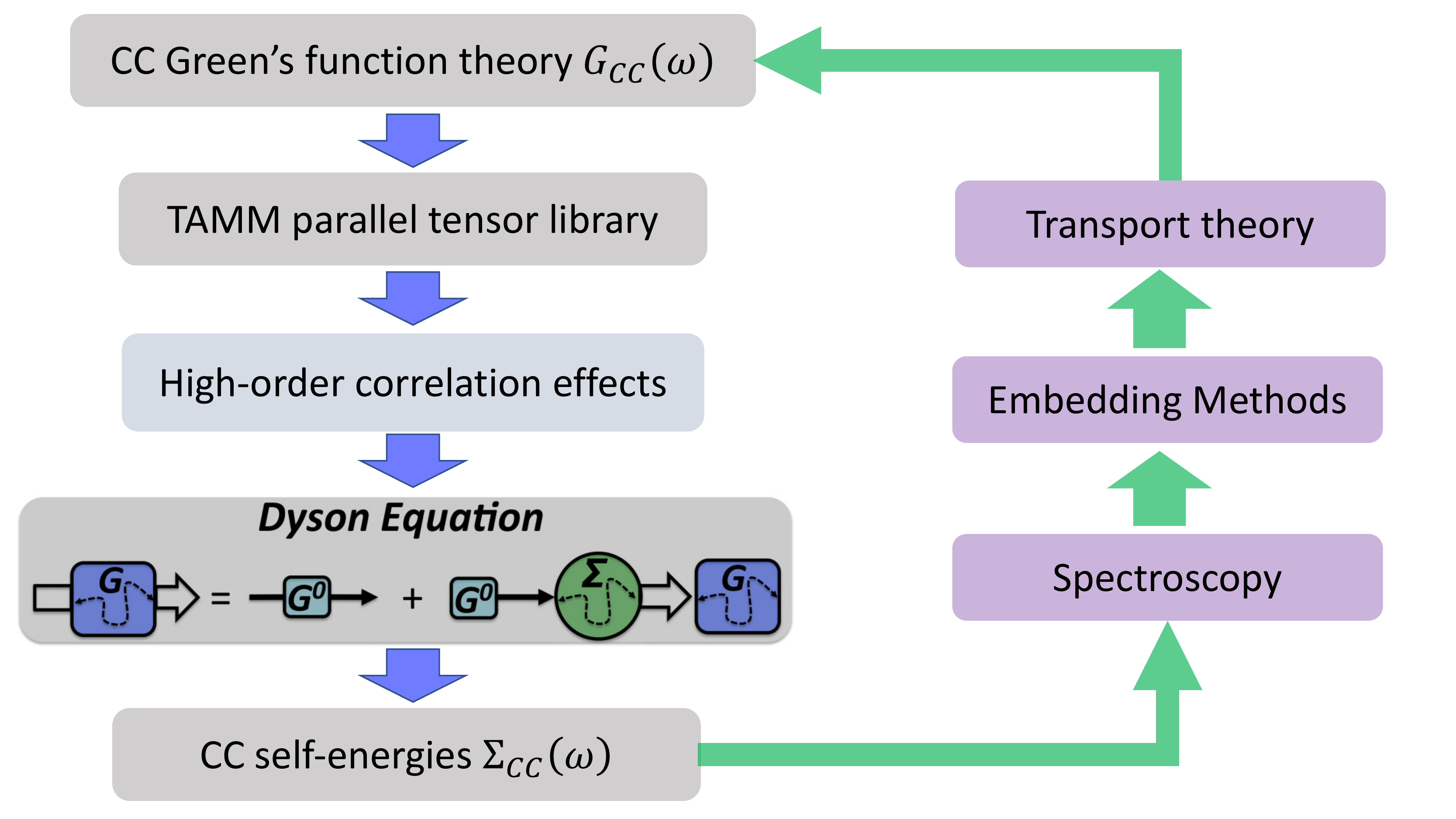}}
\caption{Schematic representation of development threads and application areas for the scalable TAMM implementations of Green's function coupled cluster.}
\label{fig_out}
\end{figure}

\section{Conclusions and Future Work}
In this paper we discussed implementation details of the GFCC approach, which utilizes capabilities of novel parallel tensor contraction library TAMM. The discussed effort integrates recent advances in  many-body quantum mechanics (GFCC theory),  applied math (MOR algorithm), and high-performance computing (TAMM library) to reduce time-to-solution associated with the construction of Green’s function matrix on leadership class GPU architectures such as OLCF Summit. The key elements of our design correspond to parallel algorithms for multi-dimensional tensors,  flexible solvers to handle large number of linear equations,  and compression algorithms  (based on the Cholesky decomposition) for class of largest 4-dimensional tensors corresponding to two-electron integrals. A special role in this effort was  played by the successful implementation of the MOR algorithm in the GFCC context, which was instrumental to enable  realistic applications of GFCC to simulate the electronic structure of large molecular systems in arbitrary frequency regime.

In the future, the TAMM infrastructure will enable quick deployment of hierarchical structure of GFCC approximations accounting for higher-rank excitations, which play an important role in identifying challenging satellite states. 
Local orbital techniques, in particular (DL)PNO, that has been utilized in the EOM-CC framework exhibiting significant reduced-scaling performance \cite{neese16_034102,neese18_244101,neese19_164123} will also be implemented in the GFCC infrastructure targeting larger and more complex systems.
These unique features make the GFCC formalism competitive to existing Green’s function approaches including family of GW and  ADC(n) formalism. We believe that these formulations can coexist and be used to construct more accurate forms of quantum embedding formalism that can utilize GFCC and through Dyson equation CC self-energies $\Sigma_{\rm CC}(\omega)$ as building blocks (see Fig.\ref{fig_out}).
A natural extension of GFCC will be core-level spectroscopy  and transport theory.

\section{Acknowledgements}
This work was supported by the Center for Scalable, Predictive methods for Excitation and Correlated phenomena (SPEC), which is funded by the U.S. Department of Energy (DOE), Office of Science, Office of Basic Energy Sciences, the Division of Chemical Sciences, Geosciences, and Biosciences.
SPEC is located at Pacific Northwest National Laboratory (PNNL)
operated for the U.S.  Department of Energy by the Battelle Memorial Institute under Contract DE-AC06-76RLO-1830.
This research used resources of the Oak Ridge Leadership Computing Facility, which is a DOE Office of Science User Facility supported under Contract DE-AC05-00OR22725.

\section{CRediT author statement}
\textbf{Bo Peng}: Investigation, Supervision, Conceptualization, Methodology, Software, Validataion, Data curation, Visualization, Writing-Original draft preparation, Reviewing and Editing;
\textbf{Ajay Panyala}: Investigation, Conceptualization, Methodology, Software, Validataion, Data curation, Visualization, Writing-Original draft preparation, Reviewing and Editing;
\textbf{Karol Kowalski}: Investigation, Supervision, Conceptualization, Methodology, Writing-Original draft preparation, Reviewing and Editing;
\textbf{Sriram Krishnamoorthy}: Investigation, Supervision, Conceptualization, Methodology, Writing-Original draft preparation, Reviewing and Editing
\bibliography{gfcc.bib}

\end{document}